\documentclass[aps,preprint,superscriptaddress]{revtex4}

\usepackage{graphicx}
\usepackage{verbatim}
\usepackage{hhline}

\newcommand{\ket}[1]{|#1\rangle}

\begin{document}

\title{Realization of the quantum Toffoli gate with trapped ions}

\author{T.~Monz}
\affiliation{Institut f\"ur Experimentalphysik, Universit\"at
Innsbruck, Technikerstr. 25, A-6020 Innsbruck, Austria}

\author{K.~Kim}
\affiliation{Institut f\"ur Experimentalphysik, Universit\"at
Innsbruck, Technikerstr. 25, A-6020 Innsbruck, Austria}

\author{W.~H\"ansel}
\affiliation{Institut f\"ur Experimentalphysik, Universit\"at
Innsbruck, Technikerstr. 25, A-6020 Innsbruck, Austria}

\author{M.~Riebe}
\affiliation{Institut f\"ur Experimentalphysik, Universit\"at
Innsbruck, Technikerstr. 25, A-6020 Innsbruck, Austria}

\author{A.~Villar}
\affiliation{Institut f\"ur Experimentalphysik, Universit\"at
Innsbruck, Technikerstr. 25, A-6020 Innsbruck, Austria}
\affiliation{Institut f\"ur Quantenoptik und Quanteninformation,
\"Osterreichische Akademie der Wissenschaften, Otto-Hittmair-Platz
1, A-6020 Innsbruck, Austria}

\author{P.~Schindler}
\affiliation{Institut f\"ur Experimentalphysik, Universit\"at
Innsbruck, Technikerstr. 25, A-6020 Innsbruck, Austria}

\author{M.~Chwalla}
\affiliation{Institut f\"ur Experimentalphysik, Universit\"at
Innsbruck, Technikerstr. 25, A-6020 Innsbruck, Austria}

\author{M.~Hennrich}
\affiliation{Institut f\"ur Experimentalphysik, Universit\"at
Innsbruck, Technikerstr. 25, A-6020 Innsbruck, Austria}

\author{R.~Blatt}
\affiliation{Institut f\"ur Experimentalphysik, Universit\"at
Innsbruck, Technikerstr. 25, A-6020 Innsbruck, Austria}
\affiliation{Institut f\"ur Quantenoptik und Quanteninformation,
\"Osterreichische Akademie der Wissenschaften, Otto-Hittmair-Platz
1, A-6020 Innsbruck, Austria}

\maketitle

{\bf Algorithms for quantum information processing are usually decomposed into sequences of quantum gate operations, most often realized with single- and two-qubit gates\cite{DiVincenzo1995PRAv51p1015}. While such operations constitute a universal set for quantum computation, gates acting on more than two qubits can simplify the implementation of complex quantum algorithms\cite{ChiaveriniNature03074}. Thus, a single three-qubit operation can replace a complex sequence of two-qubit gates, which in turn promises faster execution with potentially higher fidelity. One important three-qubit operation is the quantum Toffoli gate which performs a NOT operation on a target qubit depending on the state of two control qubits. Here we present the first experimental realization of the quantum Toffoli gate in an ion trap quantum computer. Our implementation is particular efficient as we directly encode the relevant logic information in the motion of the ion string.}


Quantum gate operations have been realized with several physical systems, e.g.\ nuclear magnetic resonance (NMR)\cite{Vandersypen2004RMPv76p1037--33}, single photons\cite{Kok2007RMPv79p135--40,Walther2005Nv434p169--176}, superconducting devices\cite{Yamamoto2003Nv425p941--944,Steffen2006Sv313p1423--1425,Plantenberg2007Nv447p836--839,MajerNature06184}, cavity QED\cite{Rauschenbeutel1999PRLv83p5166}, or ion traps\cite{Monroe1995PRLv75p4714,Leibfried2003Np412,Schmidt-Kaler2003Np408,Riebe2006PRLp220407}. All quantum gates demonstrated so far are two-qubit entangling operations, apart from one implementation of a three qubit Toffoli gate in NMR\cite{Cory1998PRLp2152}. For ion traps several schemes for implementing a quantum Toffoli gate have been proposed\cite{Chen2007TEPJD-AMOaPPv41p557--561,Ralph2007PRAv75p022313--5}. The quantum Toffoli gate is valuable in complex quantum algorithms and has an immediate practical application as correcting operation in quantum error correction\cite{Cory1998PRLp2152,SarovarProceed}. We report here on a realization of the Toffoli gate with an ion trap quantum computer which requires half the resources compared to implementations based on concatenated two-qubit gates\cite{Maslov2003ELv39p1790--1791}. This significant contraction is achieved by encoding the combined information of the two control qubits in the common vibrational mode of the ion string.

Our experimental system consists of a string of $^{40}$Ca$^{+}$ ions confined in a linear Paul trap. Each ion represents a qubit, where quantum information is stored in superpositions of the S$_{1/2}$(m=$-1/2$) = $|S\rangle \equiv |1\rangle$ ground state and the metastable D$_{5/2}$(m=$-1/2$) = $|D\rangle \equiv |0\rangle$ state of the $^{40}$Ca$^{+}$ ions\cite{Schmidt-Kaler2003APBAOp789}. The centre-of-mass (COM) vibrational mode of the ion string is used to mediate the interaction between the ion qubits. Each experiment includes a) the initialization of the qubits and the COM mode in a well defined state, b) the actual gate operation and c) a quantum state measurement. Initialization of the COM mode to the ground state is achieved by Doppler cooling on the S$_{1/2}$ $\leftrightarrow$ P$_{1/2}$ dipole transition followed by sideband cooling on the qubit transition. Optical pumping ensures that all qubits are prepared in S$_{1/2}$(m=$-1/2$). The gate operation consists of a series of laser pulses which are applied to individually addressed ions. The electronic and vibrational state of the ion string are manipulated by setting frequency, length, intensity, and phase of the pulses. The final state of the ion qubits is measured by scattering light at 397 nm on the S$_{1/2}$ $\leftrightarrow$ P$_{1/2}$ transition and detecting the fluorescence with a CCD camera. The presence or absence of fluorescence of an ion corresponds to a projective measurement in the $\{\ket{S}, \ket{D}\}$ basis. Further details about the experimental setup can be found in Ref.~\cite{Schmidt-Kaler2003APBAOp789}.

The pulse sequence for our implementation of the quantum Toffoli gate consists of three major parts: 1) Encoding of the joint quantum information of the control qubits $\ket{c_1}$ and $\ket{c_2}$ in the vibrational COM mode, 2) a NOT operation on the target qubit controlled by the vibrational mode, and 3) the reversal of the encoding step 1). Here, the encoding is the crucial part of the gate. We exploit the fact that it is always possible to add more phonons to the vibrational mode, while phonons can only be subtracted until the vibrational ground state is reached. The joint encoding of the logic control information $(c_1\ \textrm{AND}\ c_2)$ into the vibrational mode can be depicted as a two-step process: Initially, the vibrational mode contains no phonons, i.e. $\ket{vib}=\ket{n=0}$. In the first step, two phonons are deposited in the COM mode if both control ions are in $|c_{1}c_{2}\rangle=\ket{SS}$ while less than two phonons are generated for the other three basis states (see Methods). In the second step, one phonon is removed such that only for the initial state $\ket{c_1c_2}=\ket{SS}$ one phonon remains in the COM mode. Thus, we encode the logic information $(c_1\ \textrm{AND}\ c_2)$ in the vibrational mode as $(c_1\ \textrm{AND}\ c_2)=1 \Rightarrow \ket{n=1}$ and $(c_1\ \textrm{AND}\ c_2)=0\Rightarrow \ket{n=0}$. The implementation of the second step requires the use of a composite pulse sequence which realizes the subtraction of one phonon regardless of the initial phonon number (for details see Tab.~\ref{tab:pulsesequence_precise} and Methods). The next part of the Toffoli gate consists of a NOT operation on the target qubit depending on the binary information encoded in the vibrational state. This operation is provided by the central part of the Cirac-Zoller controlled NOT gate\cite{Cirac1995PRLv74p4091} as demonstrated in\cite{Schmidt-Kaler2003Np408, Riebe2006PRLp220407}. Finally, the third part reverses the encoding and restores the vibrational mode to the ground state.

The ideal unitary map implemented by our pulse sequence, see Table
\ref{tab:pulsesequence_precise}, is given by
\[  U_{\textrm{T}}=\exp(-i\pi\frac{1}{2\sqrt{2}} \sigma_{Z,t}) \left( \begin{array}{cccccccc}
    1 &   0 &   0 &   0 &   0 &   0 &   0 &   0 \\ 
    0 &   1 &   0 &   0 &   0 &   0 &   0 &   0 \\ 
    0 &   0 &   1 &   0 &   0 &   0 &   0 &   0 \\
    0 &   0 &   0 &   1 &   0 &   0 &   0 &   0 \\
    0 &   0 &   0 &   0 &   1 &   0 &   0 &   0 \\
    0 &   0 &   0 &   0 &   0 &   1 &   0 &   0 \\
    0 &   0 &   0 &   0 &   0 &   0 &   0 &   i \\
    0 &   0 &   0 &   0 &   0 &   0 &  -i &   0
\label{matrix:unitary_map}
  \end{array} \right)\]
in the basis \[ \{ | c_{1},c_{2},t\rangle \} = \{|DDD\rangle, |DDS\rangle, |DSD\rangle, |DSS\rangle, |SDD\rangle, |SDS\rangle, |SSD\rangle, |SSS\rangle \} \] The factor $\exp(-i\pi \frac{1}{2\sqrt{2}} \sigma_{Z,t})$, with the Pauli operator $\sigma_Z$, is a phase shift on the target qubit $\ket{t}$. This phase factor can either be included in a redefinition of the computational basis or compensated for by single qubit operations. In total, this unitary evolution is fully equivalent to an ideal Toffoli gate.

\begin{figure}
 \begin{center}
  \includegraphics[width=15cm]{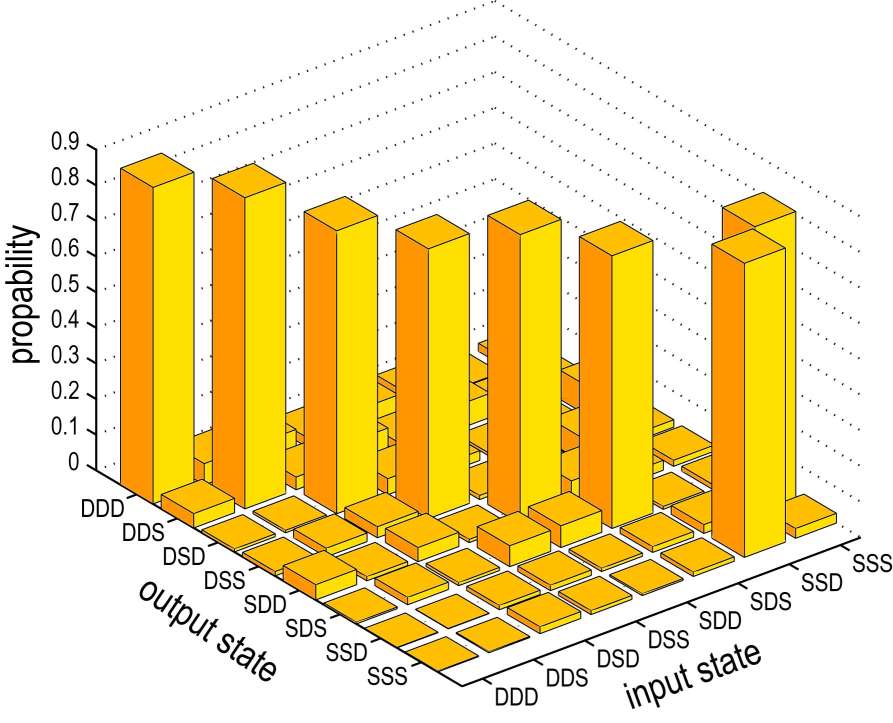}
  \caption{Experimentally obtained truth table of the Toffoli gate: After preparing the three ions in one of the eight basis input states $|SSS\rangle$ to $|DDD\rangle$ (right axis) the probabilities of all basis output states (left axis) are measured 100 times. On average the correct output state is reached with a probability of $ 81 (\pm 5) \%$. The error mainly originates from quantum projection noise.
  }
  \label{fig.truthtable}
 \end{center}
\end{figure}
The unitary map $U_{\textrm{T}}$ describes an operation which applies a NOT operation on the target ion if and only if the control ions are in the state $|SS\rangle$. We verify this behaviour by preparing the system in all eight basis states and by measuring the output state probabilities after the Toffoli operation. Figure \ref{fig.truthtable} shows a bar graph of these probabilities, the so-called truth table. Ideally, this shows the absolute values of the respective elements of the unitary matrix. On average we obtain probabilities of $\approx 81\%$ that the ion ends up in the correct output state. However, this analysis only reveals the classical features of the gate. In order to prove that we indeed perform the desired quantum operation, a more thorough analysis is required.

\begin{figure}
 \begin{center}
  \includegraphics[width=10cm]{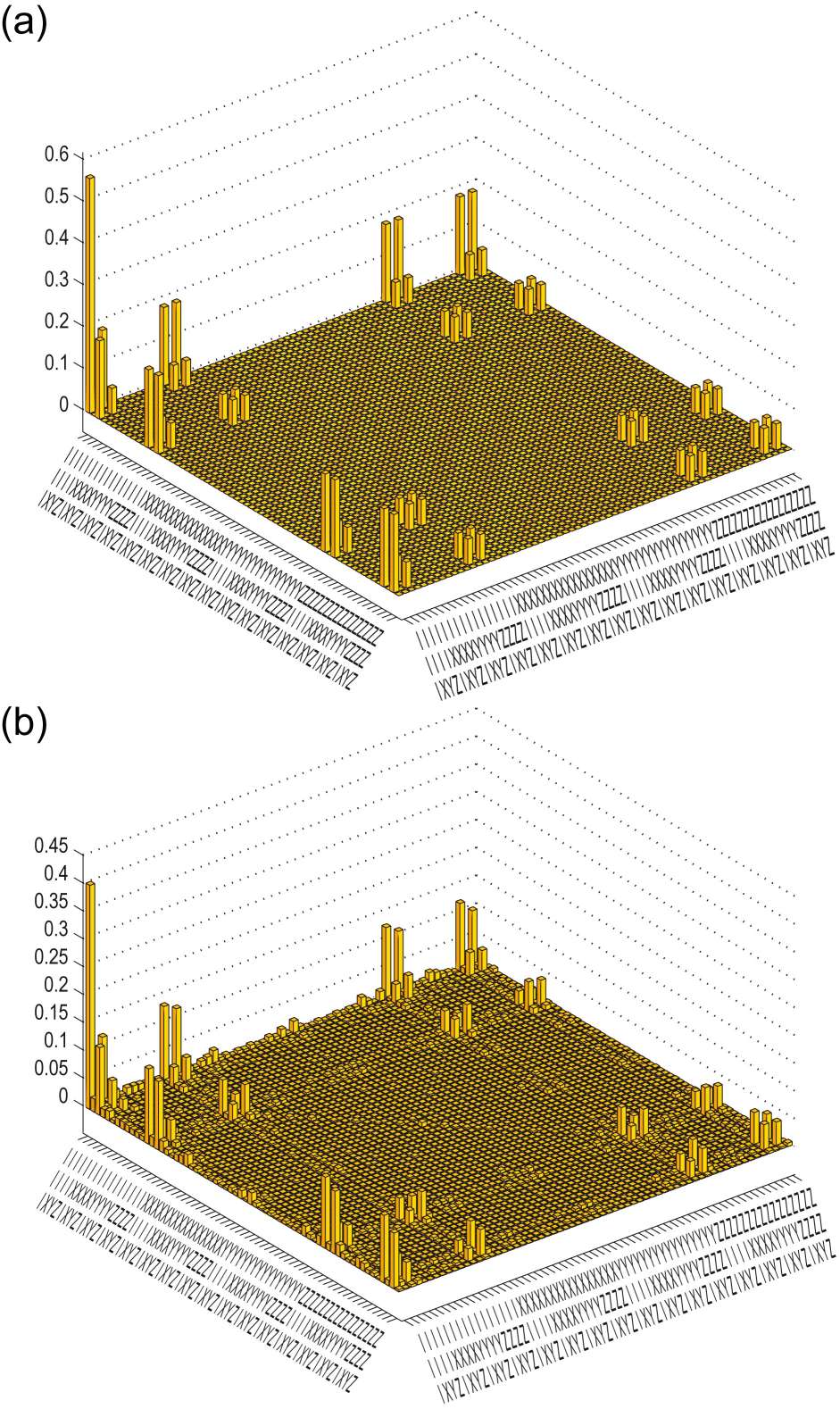}
   \caption{Comparison between (a) ideal and (b) experimental process $\chi$-matrices of the Toffoli gate. The absolute values of the elements are shown in the operator basis $\sigma_{c1}\otimes\sigma_{c2}\otimes\sigma_t \in \{I\otimes
   I\otimes I, I\otimes I\otimes X, \ldots, Z\otimes Z\otimes Z\}$. Ideal and experimental process matrices show the same ``fingerprint''.}
  \label{fig.chimatrix}
 \end{center}
\end{figure}
A complete characterization of the action of our pulse sequence must take into account experimental imperfections, in particular the coupling of our qubits to their environment, which will in general lead to a non-unitary time evolution. Such a characterization is provided by the operator sum representation in which a process $\mathcal{E}$ is described by $\rho_{out}=\mathcal{E}(\rho_{in})=\sum_{m,n=1}^{4^N} \chi_{m,n} A_{m} \rho_{in} A^{\dagger}_{n}$, where $\rho_{in}$ and $\rho_{out}$ are the input and output density matrices, $N$ is the number of qubits, and $A$ is a basis of operators in the Hilbert space of dimension $2^N \times 2^N$ \cite{chuangblackbox}. The process matrix $\chi$ contains all information about the investigated quantum operation and is obtained by quantum process tomography\cite{Nielsen2000, Riebe2006PRLp220407}. The absolute values of both the ideal and the experimentally obtained $\chi$-matrices are depicted in Fig.~\ref{fig.chimatrix}. The ``fingerprint'' of the ideal $\chi$-matrix is clearly visible in the experimental data. A quantitative measure for the performance of the gate operation is the mean gate fidelity $F_{mean}=\mathrm{mean}_{\psi_{i}}\left[\langle \psi_{i}| U_{\textrm{T}}^{\dagger}\ \mathcal{E}(|\psi_{i}\rangle \langle \psi_{i}|)\ U_{\textrm{T}} | \psi_{i} \rangle\right]$, where $U_{\textrm{T}}$ is the ideal unitary map and the $\psi_{i}$ are a set of 2 $\times$ 10$^5$ pure input states randomly drawn from the Haar measure for the unitary group $U(8)$\cite{Pozniak1998JoPAMaGv31p1059--1071}. The mean fidelity is the average value of
the fidelity between the output state expected from the ideal unitary transformation and the output state calculated from the experimentally determined $\chi$ matrix. For the results shown in Fig.~\ref{fig.chimatrix}, we obtain $F_{mean}= 71(3)\%$, where the error is due to statistical uncertainties in the tomographic measurements \cite{Roos2004PRLp220402}. The observed infidelity is mainly caused by technical imperfections. Addressing errors due to the finite width of the focused laser beam result in a weak excitation of the neighbouring ions. For the given setup, the addressing error described by the ratio of non-addressed to addressed Rabi frequency was on the order of 7\%, causing an infidelity of about 12\%. Temperature changes and voltage fluctuations of the trap electrodes produce a drift of the vibrational mode frequency. Assuming a detuning of 100 Hz, this leads to an infidelity of 7\%. Note that these two errors are purely technical imperfections and do not represent physical limitations. Remaining infidelity contributions are: initialization of the COM mode in the ground state - about 1\%; laser linewidth and magnetic field fluctuations: about 1 \%; ion state initialization: about 0.5\% per ion. The decoherence time of quantum information stored in the vibrational COM mode is on the order of 85ms, while the heating rate is on the order of 1 phonon per 140ms. These two effects do not limit the performance of the quantum Toffoli gate at the current stage.

A comparison of this result to the fidelity expected from a realization based on two-qubit CNOT gates highlights the benefits of our scheme. With our experimental setup, two-qubit CNOT gates have been demonstrated\cite{Riebe2006PRLp220407} with a mean fidelity of $F_{mean}=92.6\%$. We estimate that a realization of a composite quantum Toffoli gate with six such CNOT gates \cite{Nielsen2000} would result in a gate fidelity of approximately $F=(92.6\%)^{6} \approx 63\%$. We note that this is an optimistic estimate as it relies on two-ion data only. The duration of our gate is 1.5 ms, which is mainly determined by the coupling strength for transitions of the kind $|S,n\rangle \leftrightarrow |D,n+1\rangle$ (see Methods). A two-qubit controlled NOT gate takes approximately $700 \mu s$ \cite{Riebe2006PRLp220407}. Therefore, realizing a Toffoli gate based on six CNOT gates \cite{Nielsen2000} would be about three times longer than the sequence described above. Finally, an adaption of the proposal for the Toffoli gate using qudits \cite{Ralph2007PRAv75p022313--5} to an ion trap system would also be shorter than a decomposition into CNOT gates, but still 30\% longer than our sequence.

To conclude, we have shown the first implementation of a three-qubit Toffoli gate with trapped ions. It is notably faster than any other proposed scheme. The achieved fidelity of $F_{mean}=71(3)\%$ is higher than an optimistic estimate of the Toffoli gate fidelity based on two-qubit CNOT gates. The particular efficiency of this scheme is achieved by directly encoding the relevant logic information of the control qubits in the harmonic oscillator states of the vibrational mode. Such encoding may also help to simplify other complex quantum operations and may be applied to any system in which qubits are coupled via a harmonic oscillator. In addition, the Toffoli gate has an immediate practical application in quantum error correction. Here, depending on the state of the qubits carrying the error syndrome a conditional bit flip has to be applied to the qubits carrying the information\cite{Cory1998PRLp2152,SarovarProceed}. This task can either be accomplished by a projective measurement followed by classically controlled operations or directly by the Toffoli gate. The latter approach is particularly favourable in quantum systems where a projective measurement is not feasible or not efficient.

{\bf Methods}

  Single qubit operations are implemented by laser pulses at frequency $\omega_{0}$ resonant with the transition $|S,n\rangle \leftrightarrow |D,n\rangle$, where $n$ is the phonon number of the state. The motion of the ion string is manipulated by laser pulses at frequency $\omega_{0}+\omega_{z}$, where $\omega_{z}$ is the COM mode frequency along the trap axis, inducing transitions of the kind $|S,n\rangle \leftrightarrow |D,n+1\rangle$. This allows us to remove a phonon from the system, provided an ion is in the $|D\rangle$ state. Note that the Rabi frequency on transition $|S,n\rangle \leftrightarrow |D,n+1\rangle$ is $\Omega_{n}=\Omega_{n=0} \sqrt{n+1}$ ($\Omega_0=2\pi\ 3.3$kHz). During the encoding section, the first two pulses of the Toffoli pulse sequence (see Table \ref{tab:pulsesequence_precise}) map the state of the control qubits on the motion:
  \begin{eqnarray}
    \label{eq:1}
    |SS,0\rangle & \rightarrow & |DD,2\rangle \\
    |DS,0\rangle & \rightarrow & \sin \frac{\pi}{2\sqrt{2}} |DD,1\rangle + \cos \frac{\pi}{2\sqrt{2}} |DS,0\rangle \\
    |SD,0\rangle & \rightarrow & \cos \frac{\pi}{2\sqrt{2}} |DD,1\rangle - \sin \frac{\pi}{2\sqrt{2}} |DS,0\rangle \\
    |DD,0\rangle & \rightarrow & |DD,0\rangle
  \end{eqnarray}
  Then the first control bit is always in state $|D\rangle$. The next step is to remove exactly one phonon by transferring control ion 1 from $|D,n+1\rangle$ to $|S,n\rangle$. This is not directly possible in all four cases due to the different coupling strengths on the $|D,2\rangle \leftrightarrow |S,1\rangle$ and $|D,1\rangle \leftrightarrow |S,0\rangle$ transitions. However, a sequence of pulses of different length and phase can be tailored such that in spite of the different coupling strengths the operation $|D,n+1\rangle \leftrightarrow |S,n\rangle$ for $n=1,2$ is realized. The mechanism of this composite pulse sequence -- pulses 3--5 in Table \ref{tab:pulsesequence_precise} -- can be understood using the Bloch sphere picture with state $|D,n+1\rangle$ as the north pole and $|S,n\rangle$ as the south pole (e.g. see Ref.~\cite{Schmidt-Kaler2003APBAOp789}). Starting from $|D,2\rangle$, pulse 3 flips the state vector from the north pole to the equator, pulse 4 rotates the vector around itself, and pulse 5 continues the flip to the south pole reaching $|S,1\rangle$. On the other hand, for the initial state $|D,1\rangle$, pulse 3 rotates the state vector only partially towards the equator, pulse 4 mirrors the state vector on the equatorial plane, so that pulse 5 can complete the rotation to the south pole $|S,0\rangle$. For zero phonons nothing changes and thus conserves $|D,0\rangle$. After this step a single phonon is remaining only for the initial control state $|SS\rangle$.

\bibliography{Toffoli}

{\bf Addendum} We gratefully acknowledge support by the Austrian Science Fund (FWF), by the European Commission (SCALA, CONQUEST networks), and by the Institut f\"ur Quanteninformation GmbH. This material is based upon work supported in part by the U. S. Army Research Office.

The authors declare that they have no competing financial
interests.

Correspondence and requests for materials should be addressed to
Markus Hennrich.~(email: Markus.Hennrich@uibk.ac.at).


  \begin{table}
  \footnotesize
  \begin{tabular}{|l|l|l|l|}
    \hline
    & Pulse & Comment & Logical part\\
    \hhline{|=|=|=|=|}
    1 & $R_{1}^{+}\left(\pi,\frac{3\pi}{2}\right)$ & Encode first target qubit onto motion & Encoding \\
    \hhline{|-|-|-| |}
    2 & $R_{2}^{+}\left(\frac{\pi}{\sqrt{2}},\frac{3\pi}{2}\right)$ & Encode second target qubit onto motion & \\
    \hhline{|-|-|-| |}
    3 & $R_{1}^{+}\left(\frac{\pi}{2\sqrt{2}},\frac{\pi}{2}\right)$ & Composite pulse to remove one phonon & \\
    4 & $R_{1}^{+}\left(\pi,0\right)$ & & \\
    5 & $R_{1}^{+}\left(\frac{\pi}{2\sqrt{2}},\frac{\pi}{2}\right)$ & & \\
    \hhline{|=|=|=|=|}
    6 & $R_{3}\left(\frac{\pi}{2},0\right)$ & Prepare target qubit for motion controlled NOT & controlled NOT\\
    \hhline{|-|-|-| |}
    7 & $R_{3}^{+}\left(\frac{\pi}{2},1\right)$ & Composite phase gate & \\
    8 & $R_{3}^{+}\left(\sqrt{2}\pi,\frac{\pi}{2}\right)$ & & \\
    9 & $R_{3}^{+}\left(\frac{\pi}{2},0\right)$ & & \\
    \hhline{|-|-|-| |}
    10 & $R_{3}\left(\frac{\pi}{2},(\frac{1}{\sqrt{2}}-1)\pi\right)$ & Complete motion controlled NOT on target qubit & \\
    \hhline{|=|=|=|=|}
    11 & $R_{1}^{+}\left(\frac{\pi}{2\sqrt{2}},(-\frac{1}{2}+\frac{1}{\sqrt{2}})\pi\right)$ & Undo encoding algorithm & Decoding \\
    12 & $R_{1}^{+}\left(\pi,(-1+\frac{1}{\sqrt{2}})\pi\right)$ & & \\
    13 & $R_{1}^{+}\left(\frac{\pi}{2\sqrt{2}},(-\frac{1}{2}+\frac{1}{\sqrt{2}})\pi\right)$ & & \\
    \hhline{|-|-|-| |}
    14 & $R_{2}^{+}\left(\frac{\pi}{\sqrt{2}},(\frac{1}{2}+\frac{1}{\sqrt{2}})\pi\right)$ & Decoding finished for second control qubit & \\
    \hhline{|-|-|-| |}
    15 & $R_{1}^{+}\left(\pi,(\frac{1}{2}+\frac{1}{\sqrt{2}})\pi\right)$ & Decoding finished for first control qubit, Toffoli complete & \\
    \hline
    \end{tabular}
    \caption{ Pulse sequence of the Toffoli gate implementation: Laser pulses on the $i$th ion on the $|S,n\rangle
  \leftrightarrow |D,n\rangle$ transition are denoted by $R_{i}(\theta,\phi)$, on the $|S,n\rangle
  \leftrightarrow |D,n+1\rangle$ transition by $R^{+}_{i}(\theta,\phi)$, with laser phase $\phi$ and pulse area $\theta=\Omega t$ in terms of Rabi frequency $\Omega$, and pulse length $t$, for details see ref.~\cite{Schmidt-Kaler2003APBAOp789}.}
    \label{tab:pulsesequence_precise}
  \end{table}

\end{document}